\documentstyle[prl,aps,preprint]{revtex}
\begin{document}
\preprint{Phys.\ Rev.\ Lett.\ {\bf 72}, 1080 (1994)} 
\draft 
\title{Corrections to Scaling in the Integer Quantum Hall Effect} 
\author{Bodo Huckestein\cite{p_address}}
\address{Dept.\ of Electrical Engineering, Princeton University,
  Princeton, NJ 08544, USA} 
\date{28 September 1993} 
\maketitle
\begin{abstract}
  Finite size corrections to scaling laws in the centers of Landau
  levels are studied systematically by numerical calculations. The
  corrections can account for the apparent non-universality of the
  localization length exponent $\nu{}$. In the second lowest Landau
  level the irrelevant scaling index is
  $y_{\mathrm{irr}}=-0.38\pm0.04$. At the center of the lowest Landau
  level an additional periodic potential is found to be irrelevant
  with the same scaling index. These results suggest that the
  localization length exponent $\nu$ is universal with respect to
  Landau level index and an additional periodic potential.
\end{abstract}

\narrowtext

The transitions between different plateaus in the integer Quantum Hall
effect can be understood as disorder-driven metal-insulator
transitions in the centers of Landau levels. These transitions are
characterized by finite-size scaling laws \cite{Pru88}. Experimental
measurements of the Hall and longitudinal resistivities showed that
the corresponding localization length exponent $\nu=2.3\pm0.1$
independent of Landau level provided the spin-splitting of the levels
was resolved \cite{WTPP88,KHKP91b}. Numerically it was found that the
localization length $\lambda_M(E)$ for cylinders of circumference $M$
behaves near the critical energies $E_{\mathrm{c}}$ as
\begin{equation}
  \lambda_M(E) = M \Lambda(M/\xi(E))
  \label{fss}
\end{equation}
with $\xi(E) \propto |E-E_{\mathrm{c}}|^{-\nu}$ and $\nu=2.35\pm0.03$
\cite{CC88,HK90,Mie90,HB92,Huc92,Mie93,LD93}. This universal behavior
was observed for the lowest ($n=0$) Landau level independent of the
correlation length of the disorder potential and in the second lowest
($n=1$) Landau level provided that the correlation length was not
smaller than the magnetic length.  For shorter correlation length no
universal scaling behavior was observed and the numerical data were
inconclusive. It remained an open question whether the localization
length exponent $\nu$ was dependent on the Landau level index
\cite{Mie90,Mie93,LD93,AA85} or the available systems were too small
to observe the scaling behavior \cite{Huc92}.

Chalker and Eastmond observed that deviations from scaling behavior in
an extension of the network model with a distribution of node
parameters can be analysed in terms of irrelevant scaling fields
\cite{CE92,LWK93}. They found that the deviations of $\Lambda$ from its
fixed point value scaled like $M^{-0.38\pm0.02}$.

In this paper it is shown that their ideas can more generally explain
the observed deviations from scaling. It is found that deviations from
the finite-size scaling law Eq.~(\ref{fss}) scale by themselves and
can be described by an irrelevant scaling index $y_{\mathrm{irr}}$. In
terms of a field theory describing the transition the corrections are
due to irrelevant scaling fields \cite{Weg72,Bar83}. In particular, it
is shown that the localization length is a function of at least two
scaling fields,
\begin{equation}
  \lambda_M(E,\sigma,\dots) = M \Lambda(M/\xi(E),M/\xi_{\mathrm{irr}},
  \dots),
  \label{fss2}
\end{equation}
where $\xi_{\mathrm{irr}}$ is a function of the correlation length
$\sigma$ of the disorder potential. The function $\Lambda$ is an
analytic function of the relevant scaling field $\Delta E = (E -
E_{\mathrm{c}}) / \Gamma$ and an irrelevant scaling field
$\zeta_{\mathrm{irr}}$ that is related to the correlation length
$\sigma$. $\Gamma$ is a measure of the disorder. In the present
context the Fermi energy plays the role of the temperature in
thermodynamic phase transitions. Scaling implies that the scaling
variables are proportional to powers of the system size with the
exponents being the scaling indices
\begin{equation}
  \lambda_M(E,\sigma,\dots) = M \Lambda(M^y \Delta E,
  M^{y_{\mathrm{irr}}} \zeta_{\mathrm{irr}},\dots).
  \label{fields}
\end{equation}
$y=1/\nu$ is the only relevant, i.e. positive, scaling index,
$y_{\mathrm{irr}}$ is the largest irrelevant scaling index, and
$\dots$ represent possible further irrelevant scaling fields with
smaller scaling indices. For small arguments the function $\Lambda$
can be expanded in a Taylor series\cite{epl_note,dangerous}
\begin{equation}
  \Lambda = \Lambda_{\mathrm{c}} + a (M^y \Delta E)^2 + b
  M^{y_{\mathrm{irr}}} \zeta_{\mathrm{irr}} + \dots
  \label{lambda_series}
\end{equation}
A linear term in $\Delta E$ is missing since $\Lambda$ is symmetric in
$\Delta E$ due to the coincidence of the mobility edges at
$E=E_{\mathrm{c}}$. Eq.~(\ref{lambda_series}) is used in the following
to extract the irrelevant scaling index $y_{\mathrm{irr}}$ from the
numerical data. In the absence of any analytic information about the
scaling function $\Lambda$ it can not be ruled out that $b$ is zero
at the critical point. In this case the first non-vanishing term of
the series expansion would be quadratic in $M^{y_{\mathrm{irr}}}
\zeta_{\mathrm{irr}}$ and the numerically determined
$y_{\mathrm{irr}}$ would be twice the scaling index of the field
theory.

In order to study the corrections to scaling,
$\lambda_M(E_{\mathrm{c}})$ was calculated for $\beta^2=(\sigma^2 +
l_{\mathrm{c}}^2) / l_{\mathrm{c}}^2$, where $l_{\mathrm{c}}$ is the
magnetic length $\hbar/eB$, ranging from 1 to 2 while $M$ varied
between 16 and 128 (in multiples of $\sqrt{2\pi}l_{\mathrm{c}}$)
\cite{0.01}. For every value of $\beta^2$ the length
$\xi_{\mathrm{irr}}$ was adjusted in order to make
$\lambda_M/M(E_{\mathrm{c}})$ a function of a single variable
$M/\xi_{\mathrm{irr}}(\beta^2)$. The resulting function is shown in
Fig.~(\ref{fig:fignll1}). By performing this fit the overall scale of
$\xi_{\mathrm{irr}}$ cannot be fixed and hence is arbitrary in these
calculations. The dependence of the length scale $\xi_{\mathrm{irr}}$
on $\beta^2$ is shown in Fig.~(\ref{fig:fignll2}). It grows by more
than $10^4$ when the correlation length $\sigma$ is decreased from
$0.8 l_{\mathrm{c}}$ to 0. This large increase in the cut-off length
scale for finite-size corrections is the reason why previous
finite-size-scaling studies were unable to observe the true asymptotic
scaling behavior \cite{Huc92,Mie93,LD93}. The observation of the
corrections to scaling implies that not only the fixed point value
$\Lambda_{\mathrm{c}}$ of the scaling function but also the
localization length exponent $\nu$ are universal and independent of
Landau level index and microscopic details of the disorder. However,
in order to observe the scaling as function of $\Delta E$ the system
width $M$ would have to exceed $10^6$ for $\sigma=0$, considerably
larger than the presently accessible $M=256$ \cite{LD93}. It is not
clear why the length scale $\xi_{\mathrm{irr}}$ becomes so large in
the $n=1$ Landau level while it seems to be unnoticeably small in the
$n=0$ Landau level.

Fig.~(\ref{fig:fignll3}) shows a doubly logarithmic plot of $\Lambda -
\Lambda_{\mathrm{c}}$ as a function of $M/\xi_{\mathrm{irr}}$.
$\Lambda_{\mathrm{c}} = 1/\ln(1 + \sqrt{2}) = 1.13459\dots$ was used
which is close to the best fit estimate of $\Lambda_{\mathrm{c}} =
1.14\pm0.02$ \cite{LWK93}. The slope of the dashed line is given by
the irrelevant scaling index $y_{\mathrm{irr}}=-0.38\pm0.04$. 

Another situation where the scaling function $\Lambda(\Delta E=0)$ at
the metal-insulator-transition does not take on its critical value
$\Lambda_{\mathrm{c}}$ even for the largest numerically accessible
systems arises in the presence of a sufficiently strong additional
periodic potential \cite{Tea82,HB93,Tan93}. Here the Hamiltonian
is modified by an additional term
\begin{equation}
  V({\mathbf r}) = 4 E_0 \cos(\sqrt{2}\pi x/a)\cos(\sqrt{2}\pi y/a),
  \label{per_pot}
\end{equation}
where the period $a$ is chosen commensurable with the system width
$M$, i.e.\ $\alpha = 2\pi l_{\mathrm{c}}^2 / a^2 = q/p$, with integer
$p$ and $q$. The strength $E_0$ of the periodic potential is assumed
to be small compared to the cyclotron energy
$\hbar\omega_{\mathrm{c}}$ so that the single Landau band
approximation remains justified, but need not be small compared to the
disorder $\Gamma$. The calculations were performed for
$\delta$-correlated disorder potential in the lowest Landau level and
two different values of $\alpha$. For $\alpha=1/3$ the Landau band
splits into 3 subbands and the only critical energy is situated at the
center of the band. For $\alpha=3/5$ the Landau band splits into 5
subbands that each contain at least one critical energy for
sufficiently strong periodic potentials \cite{Tea82,HB93,Tan93}.
In both cases the energy of the critical point at the center of the
band is not changed by the periodic potential. The fitted scaling
functions $\lambda_M(E_{\mathrm{c}})/M$ are shown in
Figs.~(\ref{fig:fig3_1_1}) and (\ref{fig:fig5_3_1}) for $\alpha = 1/3$
and $3/5$, respectively.  The irrelevant length scales $\xi_{1/3}$ and
$\xi_{3/5}$ diverge approximately proportional to $E_0^m$ with $m\approx
8.7$ and $m\approx 5.8$, respectively. The irrelevant scaling indices
$y_{\alpha}$ can be deduced from Figs.~(\ref{fig:fig3_1_3}) and
(\ref{fig:fig5_3_3}) to be $y_{1/3} = -0.38\pm0.04$ and $y_{3/5} =
-0.42\pm0.04$. Based on these data there is no significant difference
between the scaling indices for $\alpha=1/3$ and $3/5$. Furthermore,
the scaling indices $y_{\mathrm{irr}}$, $y_{\alpha}$, and the one
observed by Chalker and Eastmond \cite{CE92} agree within
the numerical uncertainties \cite{Wegner}.

In conclusion, the observation of corrections to scaling according to
Eq.~(\ref{fields}) strongly supports the notion of {\em universal\/}
metal-insulator-transitions at the centers of Landau levels in the
integer Quantum Hall Effect. The occurrence of very large irrelevant
length scales explains why the universality of the localization length
exponent $\nu$ could not be observed directly in previous
calculations.  According to an argument by Lee, Wang and Kivelson the
scaling function $\Lambda$ is related to the longitudinal conductivity
$\sigma_{xx}$ \cite{LWK93}. The universal value of
$\Lambda_{\mathrm{c}}$ would thus imply that the peak value of
$\sigma_{xx}$ in the center of each Landau level is 1/2 $e^2/h$,
independent of Landau level index. It is further shown that an
additional periodic potential is an irrelevant perturbation at the
critical point even though it can create additional critical states in
each Landau level \cite{Tea82,HB93,Tan93}.  The observed values for
the largest irrelevant scaling indices,
$y_{\mathrm{irr}}=-0.38\pm0.04$, $y_{1/3}=-0.38\pm0.04$, and
$y_{3/5}=-0.42\pm0.04$, are further important parameters, besides the
localization length exponent $\nu=2.35\pm0.03$ \cite{Huc92}, that
could be used to check an analytic theory of the integer Quantum Hall
effect.

I gratefully acknowledge the support of a DFG scholarship, as well as
the hospitality of the Aspen Center for Physics where this work was
started. I thank J.T. Chalker for telling me about the results of
Ref.\cite{CE92} before publication. I benefited greatly from
discussions with him, M.P.A.  Fisher, S.M. Girvin, and A.W.W. Ludwig.

%\bibliographystyle{prsty}
%\bibliography{bodos,qhe,current}

\begin{thebibliography}{10}
\bibitem{epl_note}
In Ref.~\cite{Huc92} it was claimed that $\Lambda$ can be well approximated by
  $\Lambda = \Lambda_{\mathrm{c}} + a M^y |\Delta E|$ with
  $\Lambda_{\mathrm{c}}=1.19\pm0.04$. However, a careful reexamination
  of the data presented in Ref.~\cite{Huc92} shows that a quadratic
  $\Delta E$ dependence, necessary for analyticity of $\Lambda{}$ in
  $\Delta{}E$, and $\Lambda_{\mathrm{c}}=1.13\pm0.01$ fits
  the data better for small $\Delta{}E$.
\bibitem{dangerous}
It is assumed that $\zeta_{\mathrm{irr}}$ is not a dangerous
  irrelevant variable.
\bibitem{0.01}
Actually, the calculations were performed for $\Delta E=0.01$. However, due to
  the quadratic dependence of $\Lambda$ on $\Delta E$, the resulting error is
  much smaller than the statistical error.
\bibitem{Wegner}
It seems worthwhile to note that Wegner's correction-to-scaling exponent
  $\Delta_i=|y_{\mathrm{irr}}|/y=|y_{\mathrm{irr}}|\nu$ turns out to be
  $0.94\pm0.1$ \cite{Weg72}. Since this is compatible with
  $\Delta_i=1$ it might imply that corrections to scaling are analytic
  in $\Delta E$.
\end{thebibliography}

\begin{figure}
\caption[fignll1]{\label{fig:fignll1} The renormalized exponential decay length
  $\lambda_{M}/M$ for $\beta^2=1.0$ ($\circ$), 1.1 ($\diamond$), 1.2
  ($\ast$), 1.3 ($\star$), 1.4 ($\times$), 1.5 ($\bullet$), 1.6
  (${\scriptstyle\bigtriangleup}$), and 1.7
  (${\scriptstyle\bigtriangledown}$). The dashed line represents the
  asymptotic value $\Lambda_{\mathrm{c}}=1/\ln(1+\sqrt{2})$.}
\end{figure}
\begin{figure}
\caption[fignll2]{\label{fig:fignll2} The length scale
  $\xi_{\mathrm{irr}}$ (in units of $\sqrt{2\pi}l_{\mathrm{c}}$) as a
  function of $\beta^2$. The value at $\beta^2=1$ has been arbitrarily
  fixed to $\xi_{\mathrm{irr}}=50\,000$.}
\end{figure}
\begin{figure}
\caption[fignll3]{\label{fig:fignll3} Deviation of the scaling
  function $\Lambda$ from its asymptotic value $\Lambda_{\mathrm{c}}$.
  The data and fitted scaling function of Fig.~(\ref{fig:fignll1}) are
  shown. The scatter of the data for large $M/\xi_{\mathrm{irr}}$ is
  due to the statistical errors of the data that become comparable to
  the deviation $\Lambda-\Lambda_{\mathrm{c}}$. The dashed (shifted)
  line with slope $y_{\mathrm{irr}}=-0.38$ serves as a guide to the
  eye.}
\end{figure}
\begin{figure}
\caption[fig3_1_1]{\label{fig:fig3_1_1} The renormalized exponential
  decay length $\lambda_{M}/M$ for $\alpha=1/3$ and $E_0=0.75$
  ($\circ$), 1 ($\diamond$), 1.25 ($\ast$), and 1.5 ($\star$). The
  dashed line represents the asymptotic value $\Lambda_{\mathrm{c}}$.}
\end{figure}
\begin{figure}
\caption[fig5_3_1]{\label{fig:fig5_3_1} The renormalized exponential
  decay length $\lambda_{M}/M$ for $\alpha=3/5$ and $E_0=1$ ($\circ$),
  1.25 ($\diamond$), 1.5 ($\ast$), 1.75 ($\star$), 2 ($\times$), and
  2.25 ($\bullet$). The dashed line represents the asymptotic value
  $\Lambda_{\mathrm{c}}$.}
\end{figure}
\begin{figure}
\caption[fig3_1_3]{\label{fig:fig3_1_3} Deviation of the scaling
  function $\Lambda$ from its critical value
  $\Lambda_{\mathrm{c}}=1/\ln(1+\sqrt{2})$ for $\alpha=1/3$. The
  dashed line has slope $y_{1/3}=-0.38$ (cf.\ 
  Fig.~(\ref{fig:fignll3})).}
\end{figure}
\begin{figure}
\caption[fig5_3_3]{\label{fig:fig5_3_3} Deviation of the scaling
  function $\Lambda$ from its critical value
  $\Lambda_{\mathrm{c}}=1/\ln(1+\sqrt{2})$ for $\alpha=3/5$. The
  dashed line has slope $y_{3/5}=-0.42$ (cf.\ 
  Fig.~(\ref{fig:fignll3})).}
\end{figure}
\end{document}